\documentclass[a4paper]{jpconf}
\usepackage{graphicx}
\begin{document}
\title{Reproducible Experiment Platform}

\author{Tatiana Likhomanenko$^{1,2,3}$, Alex Rogozhnikov$^{1,2}$, Alexander Baranov$^{1}$, Egor Khairullin$^{1,4 }$, Andrey Ustyuzhanin$^{1,2,3,4}$}

%\address{$^1$ Mathematics Faculty, Open University, 
%Milton Keynes MK7~6AA, UK}
%\address{$^2$ Department of Mathematics, 
%Imperial College, Prince Consort Road, London SW7~2BZ, UK}
%\address{$^3$ Department of Computer Science, 
%University College London, Gower Street, London WC1E~6BT, UK}

\address{$^1$ Yandex School of Data Analysis (YSDA), RU}
\address{$^2$ National Research University Higher School of Economics (HSE), RU}
\address{$^3$ NRC "Kurchatov Institute", RU}
\address{$^4$ Moscow Institute of Physics and Technology, Moscow, RU}

\ead{antares@yandex-team.ru, axelr@yandex-team.ru}

\begin{abstract}
Data analysis in fundamental sciences nowadays is an essential process that pushes frontiers of our knowledge and leads to new discoveries. At the same time we can see that complexity of those analyses increases fast due to a)~enormous volumes of datasets being analyzed, b)~variety of techniques and algorithms one have to check inside a single analysis, c)~distributed nature of research teams that requires special communication media for knowledge and information exchange between individual researchers. There is a lot of resemblance between techniques and problems arising in the areas of industrial information retrieval and particle physics. To address those problems we propose Reproducible Experiment Platform (REP), a software infrastructure to support collaborative ecosystem for computational science. It is a Python based solution for research teams that allows running computational experiments on shared datasets, obtaining repeatable results, and consistent comparisons of the obtained results. We present some key features of REP based on case studies which include trigger optimization and physics analysis studies at the LHCb experiment.
\end{abstract}

\section{Introduction}
This paper presents Reproducible Experiment Platform (REP,~\cite{REP_cite}), which is created to perform a reproducible data analysis in a comfortable way. As Karl Popper, one of the greatest philosophers of science, said: "Non-reproducible single occurrences are of no significance to science". 

Being a significant part of research, scientific analyses should be prepared in a reproducible way. However, there are several reasons why many of analyses prepared in recent years have difficulties in reevaluation: keeping data, code and results together; independence on development platform; availability of different algorithms; distributed research; repeatable data preprocessing. Toolkit presented here, REP, addresses these problems (will be discussed further) and, moreover, makes it possible to  collaborate by sharing analysis code and results over the Internet. This platform provides methods for preparing and processing data, ways to use different machine learning techniques. Yet, we are hoping to have readable and, which is important, checkable code using the platform.

\section{REP Infrastructure}

\begin{figure}
\includegraphics[width=38pc]{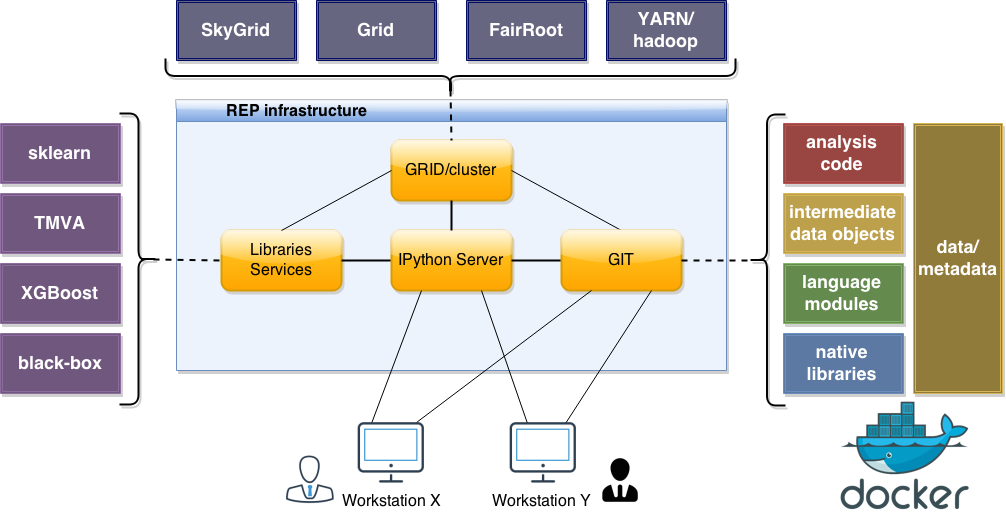}\hspace{2pc}%
\caption{\label{rep} REP infrastructure. IPython notebook is an analysis environment, in which users work together. In the notebook  we use different machine learning libraries (left) for data analysis, parallel systems to speedup optimal model search (top), {\tt git} to store all code, data and results (right).}
\end{figure}

What are the main points we expect from an analysis platform?
Firstly, we need some interactive environment for fast experiments with data. Secondly, resulting code should be simple and reproducible. Thirdly, it is necessary to keep data, code and results together. REP covers all three features.

Interactive Python, or IPython notebook, is used as an analysis environment (see Figure~\ref{rep}) in REP. For intellectual data analysis we use different machine learning libraries (apart from TMVA), which are popular among data scientists. For this purpose REP contains wrappers over algorithms from different machine learning libraries, which provide {\tt scikit-learn} interface~\cite{sklearn}. A typical analysis task, search for optimal predictive model, can be speeded up using parallel computational system. In Figure~\ref{rep} they are presented as a GRID/cluster system. One of the parallel execution systems is provided out-of-the-box by IPython (IPython cluster). Another significant part of experiments is {\tt git}~\cite{git}, version control system, which stores all code, corresponding results, trained models and data.
 
One of the most significant problems for reproducibility is keeping track of versions of all libraries used (and their numerous dependencies). This can be archived by using virtual machine, where a scientist saves his analysis with all dependencies. However, a better option exists today: one can use like a light-weight virtual machine, Docker container~\cite{docker}. The images of virtual machines can be combined together inside a Docker to provide possibilities given by different containers. It has several other advantages, 
among which incremental versioning of containers. This versioning implies that to change version of container user doesn't need to reload complete image, only some `update' part is downloaded. That is why we provide a Docker container with REP and all it dependencies. Being a virtual machine, the REP container is expected to work after many years on any hardware and operating system.

\begin{figure}
\begin{minipage}{36pc}
\includegraphics[width=38pc]{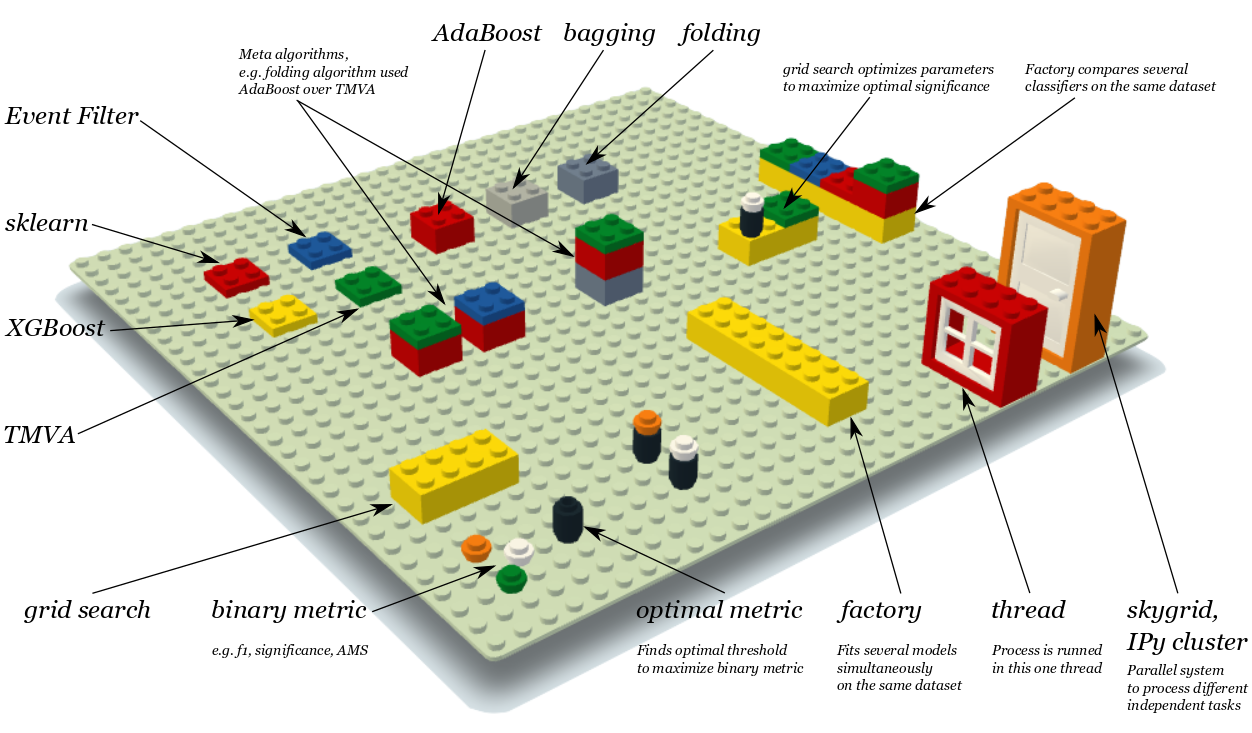}
\caption{\label{rep_basic} REP basic elements for constructing complicated analysis scheme.}
\end{minipage}\hspace{2pc}%
\end{figure}

\section{Machine Learning Pipeline}
The main feature of REP is a simultaneous support of different machine learning libraries. Wrappers over all libraries use scikit-learn interface because of its popularity in data science community and convenience for machine learning purposes. Moreover, there are other advantages: support of scikit-learn interface makes it possible to use ensembling algorithms from this library. For example, REP provides TMVA~\cite{tmva} wrapper in scikit-learn interface and ensembling algorithm over any TMVA method can be constructed. One can construct scikit-learn AdaBoost over TMVA rectangular cut method, or TMVA multilayer perceptron, or any another TMVA method. This way of combining different methods is very typical for scikit-learn.

Wrappers over machine learning libraries are basic elements to construct complicated analysis scheme using ensembling algorithms or another hierarchical type models. REP machine learning main elements are described in Figure~\ref{rep_basic}. Fundamental elements are wrappers over libraries: TMVA, scikit-learn, XGBoost~\cite{xgb}, Event Filter, \textit{etc}. Event Filter is a web-service for machine learning provided by Yandex, which is available only for CERN members. Event Filter uses MatrixNet algorithm~\cite{mn_paper} developed at Yandex.

REP constructs ensembling and hierarchical models using basic elements: scikit-learn AdaBoost ensemble algorithm, bagging, folding (used in high energy physics data analyses),~\textit{etc}. Any wrapper can be base estimator for any of these hierarchical models. It is necessary to find the best model among of all these configurations. For this purpose REP provides a grid search, which takes any estimator and looks for the best estimator's hyper parameters by optimizing quality metric selected. The grid search can optimize parameters for all hierarchical levels of complicated estimator (for instance, this allows tuning number of stages in AdaBoost and parameters of base classifier used by AdaBoost). One more task frequently arising is different models training and comparison of  their performance. Specially for these purposes REP contains a factory. To speed up training operations parallel system (IPython cluster, threads) is available during factory and grid search fitting.

\begin{figure}
\begin{minipage}{38pc}
\includegraphics[width=38pc]{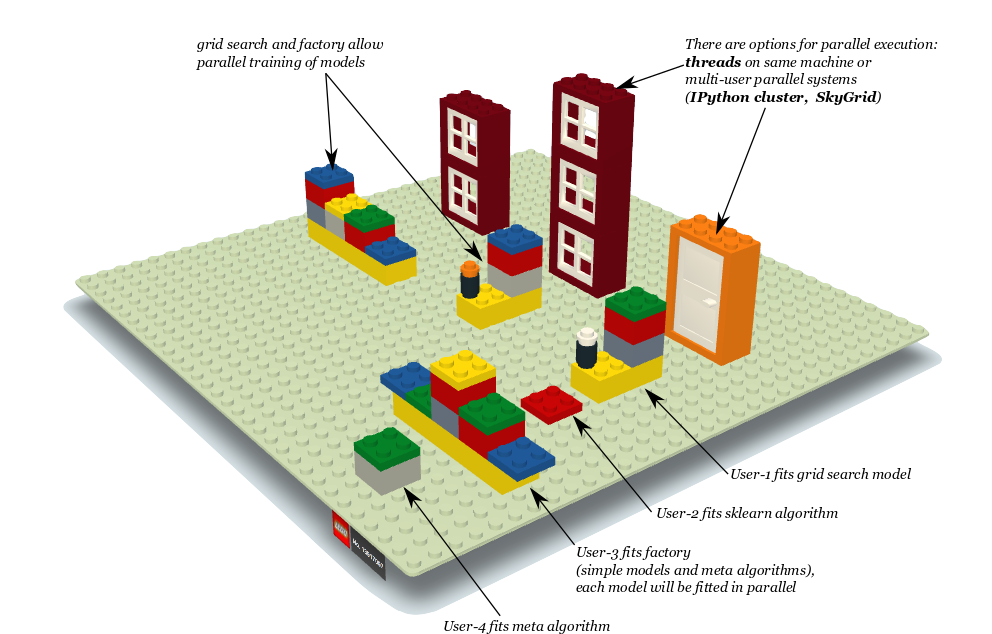}
\caption{\label{rep_comp} Examples of possible complicated data analysis schemes: factory training in the two threads, grid search training in the three threads, different models training in the multi users parallel system.}
\end{minipage} 
\end{figure}

\begin{figure}
\begin{minipage}{18pc}
\includegraphics[width=18pc]{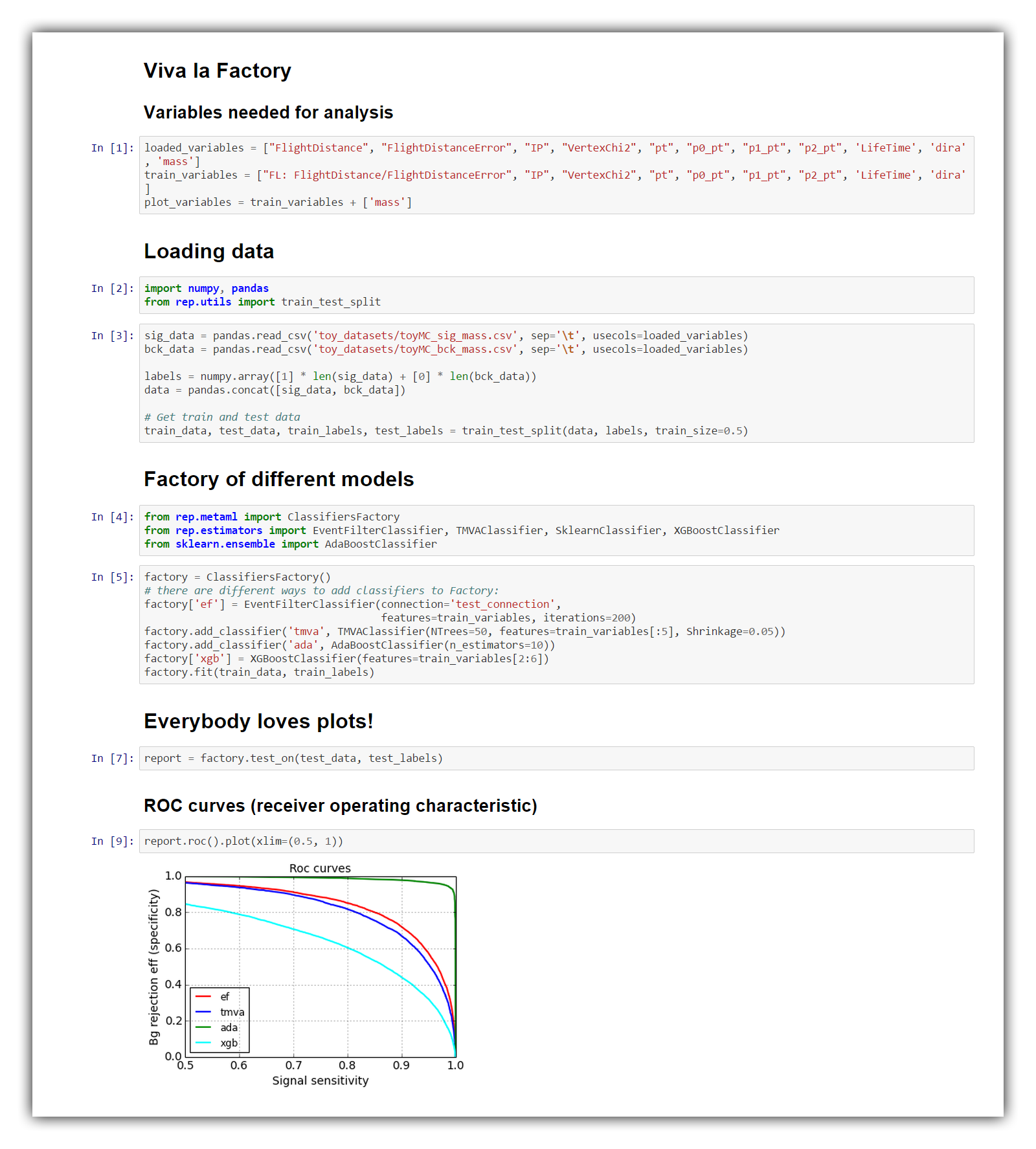}
\caption{\label{ex1} An example of using a factory for training and comparing classifiers.}
\includegraphics[width=18pc]{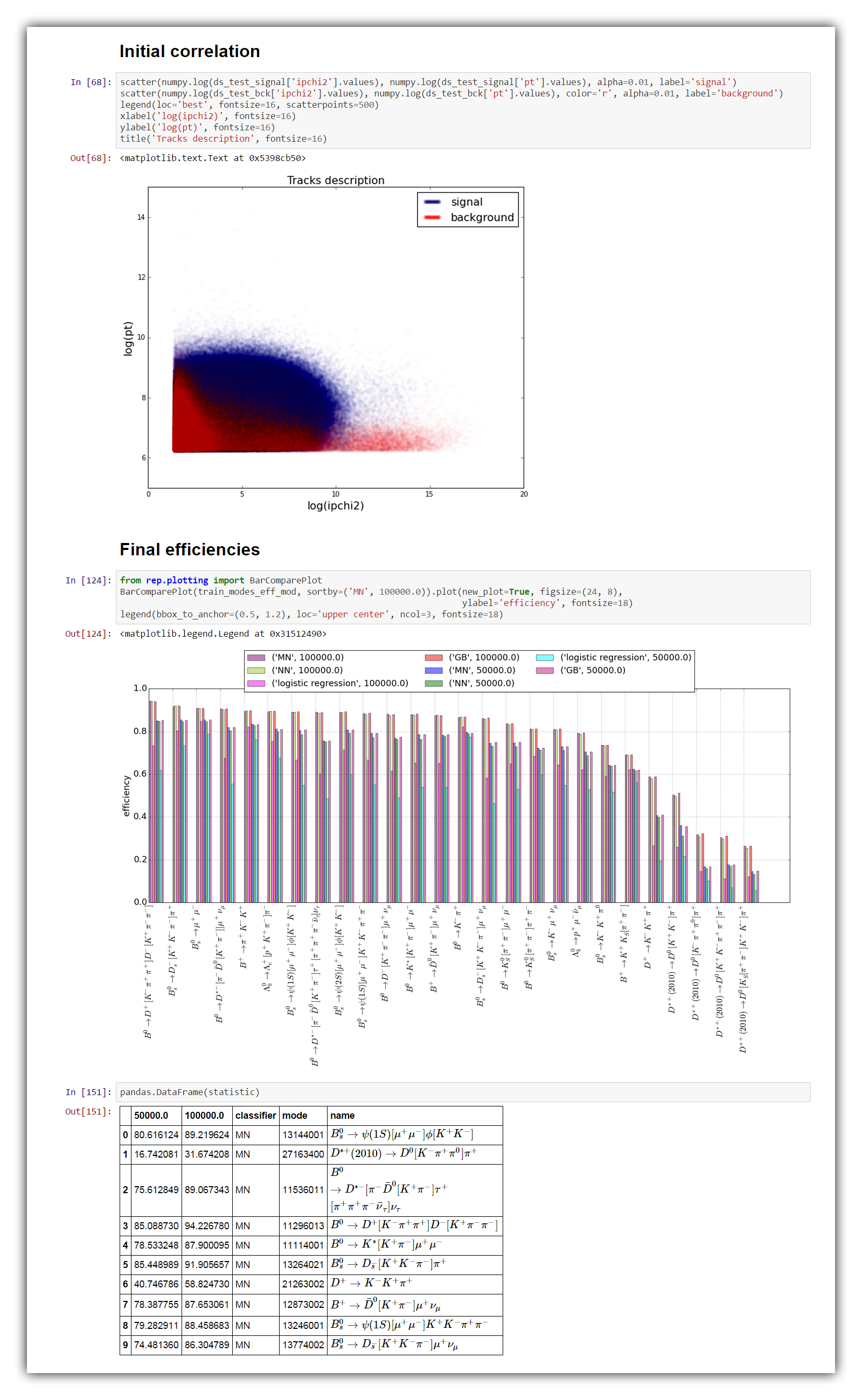}
\caption{\label{ex2} Case-study: part of LHCb topological trigger reoptimization notebook.}
\end{minipage} \hspace{2pc}%
\begin{minipage}{18pc}
\includegraphics[width=18pc]{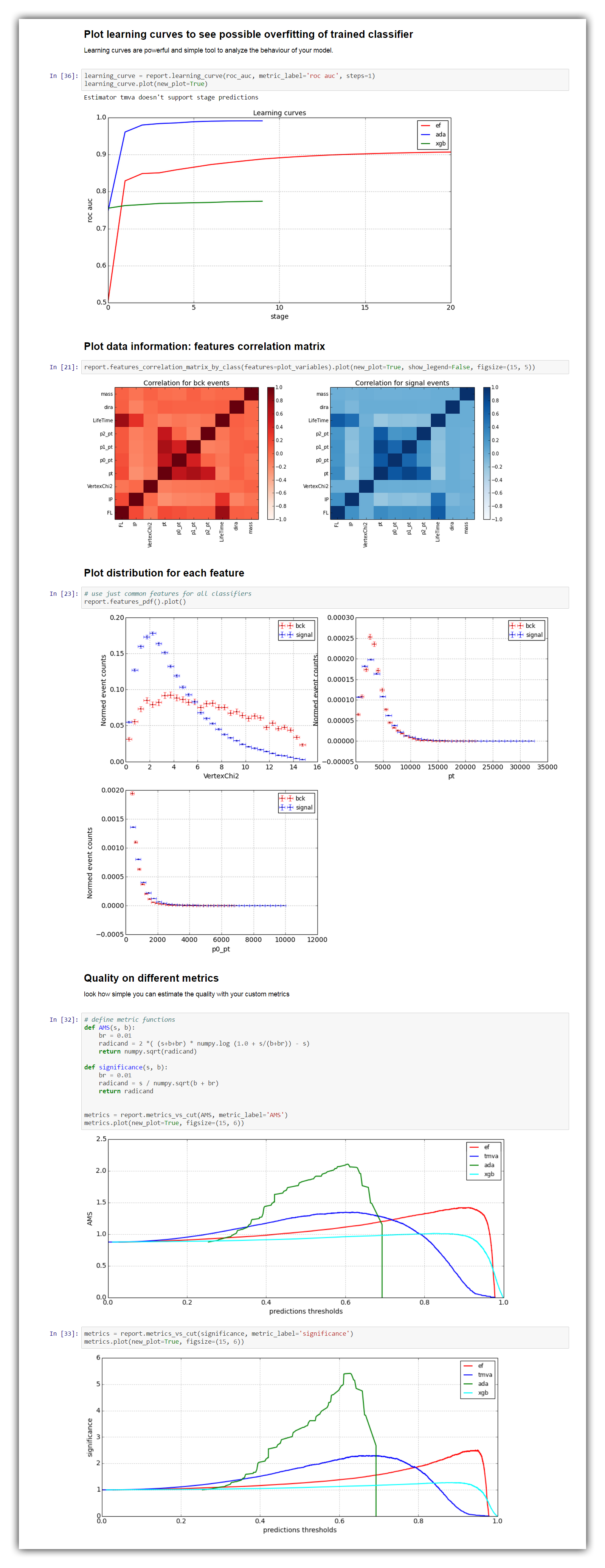}
\caption{\label{ex3} An example of comparing models and plotting results with REP.}
\end{minipage}
\end{figure}

Thus it is possible to construct machine learning schemes like those shown in Figure~\ref{rep_comp}. Examples of data analysis notebooks look like Figures~\ref{ex1},~\ref{ex2},~\ref{ex3}.

\section{IPython Keeper Extension}
One of the REP part is an IPython keeper, or IPykee, which provides comfortable interface to create project and add notebooks with code and saved results: plots, estimators, data, another objects. All of this is saved to the repository of {\tt git} version control system. Later anyone can load project and reproduce experiment or load its results at particular point in research. Also, by using nbdiff library, IPykee gives possibility for visual comparing between two versions of one notebook with marked differences.

\section{Conclusion}
To sum up, Reproducible Experiment Platform, REP, provides the environment to conduct reproducible data analysis in a convenient way. It combines different machine learning libraries under uniform interface, meta-algorithms, different parallel systems and contains tools to save intermediate results and states of analysis. REP was used in several public projects, such as Data popularity  ({http://github.com/hushchyn-mikhail/DataPopularity}), uniforming (uBoost-like) algorithms \cite{uboost} ({http://github.com/anaderi/lhcb\_trigger\_ml}), LHCb topological trigger optimization. Its use case is currently expanded to physics analyses of LHCb.

\end{document}